\documentclass[11pt]{article}
\pdfoutput=1
\usepackage{epsfig,cite,color,amsmath,amssymb}
\newcommand{\gev}{{\rm GeV}}

\def\nn{\nonumber}
\def\be{\begin{equation}}
\def\ee{\end{equation}}
\newcommand{\bea}{\begin{eqnarray}}
\newcommand{\eea}{\end{eqnarray}}
\newcommand{\bdm}{\begin{displaymath}}
\newcommand{\edm}{\end{displaymath}}
\long\def\symbolfootnote[#1]#2{\begingroup%
\def\thefootnote{\fnsymbol{footnote}}\footnote[#1]{#2}\endgroup}

\def\citeloro{[33]}
\def\fhversion{2.10.2}

\def\sq2{\sqrt{2}}
\def\drbar{\overline{\rm DR}}

\def\smallOS{\scriptscriptstyle{\rm OS}}
\def\smalldr{\scriptscriptstyle{\overline{\rm DR}}}

\def\tb{\tan\beta}
\def\cbe{c_\beta}
\def\sbe{s_\beta}

\def\cbeq{c^2_\beta}
\def\sbeq{s^2_\beta}
\def\gl{\tilde{g}}
\def\mg{m_{\gl}}
\def\g{\mg^2}

\def\x1g{x_{1}}

\def\mhmax{m_h^{\rm max}}

\newcommand{\as}{\alpha_s}
\newcommand{\oas}{{\cal O}(\as)}
\newcommand{\smallz}{{\scriptscriptstyle Z}} %  a smaller Z
\newcommand{\smallw}{{\scriptscriptstyle W}} %
\newcommand{\smallH}{{\scriptscriptstyle H}} %
\newcommand{\smallr}{{\scriptscriptstyle R}} %
\newcommand{\smalla}{{\scriptscriptstyle A}} %
\newcommand{\mz}{m_\smallz}
\newcommand{\mw}{m_\smallw}
\newcommand{\mh}{m_h}
\newcommand{\ma}{m_\smalla}
\newcommand{\mH}{m_\smallH}
\newcommand{\mhb}{\overline{m}_h}
\newcommand{\mHb}{\overline{m}_\smallH}
\newcommand{\alb}{\overline{\alpha}}
\newcommand{\muR}{\mu_\smallr}

\def\at{\alpha_t}

\def\ab{\alpha_b}
\def\as{\alpha_s}
\def\oas{{\cal O}(\as)}
\def\oat{{\cal O}(\at)}

\def\oaas{{\cal O}(\alpha\as)}
\def\oatas{{\cal O}(\at\as)}
\def\oabas{{\cal O}(\ab\as)}
\def\oatababq{{\cal O}(\at\ab + \ab^2)}

\def\oatq{{\cal O}(\at^2)}

\def\mt{m_t}
\def\stu{\tilde{t}_1}
\def\std{\tilde{t}_2}

\def\t{\mt^2}

\def\tu{m_{\tilde{t}_1}^2}
\def\td{m_{\tilde{t}_2}^2}

\def\sdt{s_{2\theta_t}}

\begin{document}

\begin{titlepage}

%\today

{\flushright{
        \begin{minipage}{2.7cm}
          RM3-TH/14-15\\MPP-2014-373
        \end{minipage}        }

}
\renewcommand{\thefootnote}{\fnsymbol{footnote}}
\vskip 2cm
\begin{center}
\boldmath
{\LARGE\bf Two-loop QCD corrections to the MSSM Higgs masses}
\vskip 0.3cm
{\LARGE\bf beyond the effective-potential approximation}\unboldmath
\vskip 1.cm
{\Large{G.~Degrassi$^{a}$, S.~Di~Vita$^{b}$ and P.~Slavich$^{c,d}$}}
\vspace*{8mm} \\
{\sl ${}^a$
    Dipartimento di Matematica e Fisica, Universit\`a di Roma Tre and  INFN, 
    Sezione di Roma Tre \\
    Via della Vasca Navale~84, I-00146 Rome, Italy}
\vspace*{2mm}\\
{\sl ${}^b$ Max-Planck-Institut f\"ur Physik, F\"ohringer Ring 6, D-80805 Munich, Germany}
\vspace*{2mm}\\
{\sl ${}^c$ LPTHE, UPMC Univ.~Paris 06,
  Sorbonne Universit\'es, 4 Place Jussieu, F-75252 Paris, France}
\vspace*{2mm}\\
{\sl ${}^d$ LPTHE, CNRS, 4 Place Jussieu, F-75252 Paris, France }
\end{center}
\symbolfootnote[0]{{\tt e-mail:}}
\symbolfootnote[0]{{\tt degrassi@fis.uniroma3.it}}
\symbolfootnote[0]{{\tt divita@mpp.mpg.de}}
\symbolfootnote[0]{{\tt slavich@lpthe.jussieu.fr}}

\vskip 0.7cm

\begin{abstract}
We compute the two-loop QCD corrections to the neutral Higgs-boson
masses in the MSSM, including the effect of non-vanishing external
momenta in the self-energies.  We obtain corrections of $\oatas$ and
$\oaas$, i.e., all two-loop corrections that involve the strong gauge
coupling when the only non-vanishing Yukawa coupling is the top
one. We adopt either the $\drbar$ renormalization scheme or a mixed
OS--$\drbar$ scheme where the top/stop parameters are renormalized
on-shell. We compare our results with those of earlier calculations,
pointing out an inconsistency in a recent result obtained in the 
mixed OS--$\drbar$ scheme. The numerical impact of
the new corrections on the prediction for the lightest-scalar mass is
moderate, but already comparable to the accuracy of the Higgs-mass
measurement at the LHC.
\end{abstract}
\vfill
\end{titlepage}    
\setcounter{footnote}{0}

\section{Introduction}
\label{sec:intro}

The accuracy of the measurement of the Higgs-boson mass by the ATLAS
and CMS collaborations at the Large Hadron Collider (LHC) has already
reached the level of $300\!-\!400$~MeV~\cite{ATLASmass,CMSmass} and,
being still dominated by statistics, is bound to improve further when
the LHC restarts operations in 2015. This puts new emphasis on the
need for high-precision calculations in those extensions of the
Standard Model (SM), such as the Minimal Supersymmetric Standard Model
(MSSM), in which the Higgs-boson mass can be predicted as a function
of other physical observables.

The Higgs sector of the MSSM consists of two $SU(2)$ doublets, $H_1$
and $H_2$, whose relative contribution to electroweak (EW) symmetry
breaking is determined by the ratio of vacuum expectation values
(VEVs) of their neutral components, $\tb\equiv v_2/v_1$. The spectrum
of physical Higgs bosons is richer than in the SM, consisting of two
neutral scalars, $h$ and $H$, one neutral pseudoscalar, $A$, and two
charged scalars, $H^\pm$. At the tree level, the neutral scalar masses
$\mh$ and $\mH$ and the scalar mixing angle $\alpha$ can be computed
in terms of the $Z$-boson mass $\mz$, the pseudoscalar mass $\ma$ and
$\tb$, and the bound $\mh < |\cos2\beta|\,\mz$ applies.  In a
significant portion of the parameter space the lightest scalar $h$ has
SM-like couplings to fermions and gauge bosons, in which case the
tree-level bound on $\mh$ has long been disproved by the
LEP~\cite{LEPHiggs}.  However, radiative corrections can raise the
prediction for the lightest-scalar mass up to the value $\mh \approx
125$~GeV observed at the LHC, and they bring along a dependence on all
MSSM parameters. Among the latter, particularly relevant are the
masses and mixing of the scalar partners of the third-generation
quarks, the stop and sbottom squarks.

Due to the crucial role of radiative corrections in pushing the
prediction for the lightest-scalar mass above the tree-level bound, an
impressive theoretical effort has been devoted over more than twenty
years to the precise determination of the Higgs sector of the
MSSM.\footnote{\,We focus here on the MSSM with real
  parameters. Significant efforts have also been devoted to the
  Higgs-mass calculation in the presence of CP-violating phases, as
  well as in non-minimal supersymmetric extensions of the SM.} After
the early realization~\cite{higgsearly} of the importance of the
one-loop $\oat$ corrections\,\footnote{\,We define $\alpha_{t,b} =
  h_{t,b}^2/(4\pi)$, where $h_t$ and $h_b$ are the superpotential top
  and bottom couplings, respectively. We follow the standard
  convention of denoting by $\oat$ the one-loop corrections to the
  Higgs masses that are in fact proportional to $h_t^2 m_t^2$, i.e.~to
  $h_t^4 v_2^2$. Similar abuses of notation affect the other one- and
  two-loop corrections.}  involving top and stop, full one-loop
computations of the MSSM Higgs masses have been
provided~\cite{brignole,1loop}, leading logarithmic effects at two
loops have been included via renormalization-group methods~\cite{rge},
and genuine two-loop corrections of
$\oatas$~\cite{hemphoang,Sven,Zhang,ezhat,dsz},
$\oatq$~\cite{hemphoang,ezhat,bdsz}, $\oabas$ \cite{bdsz2,heidi} and
$\oatababq$~\cite{dds} have been evaluated in the limit of vanishing
external momentum in the Higgs self-energies. All of these corrections
have been implemented in widely-used computer codes for the
calculation of the MSSM mass spectrum, such as {\tt
  FeynHiggs}~\cite{feynhiggs}, {\tt SoftSUSY}~\cite{softsusy}, {\tt
  SuSpect}~\cite{suspect} and {\tt SPheno}~\cite{spheno}.
Furthermore, a complete two-loop calculation of the MSSM Higgs masses
in the effective potential approach (i.e., at zero external momentum),
including also two-loop corrections controlled by the EW gauge
couplings, has been presented in ref.~\cite{martin_effpot}.  Some of
the dominant three-loop corrections to $\mh$ have also been obtained,
both via renormalization-group methods~\cite{3loopMartin,heavystops}
and by explicit calculation of the Higgs self-energy at zero external
momentum~\cite{3loopFD}.

Already at the two-loop level, going beyond the approximation of zero
external momentum brings significant complications to the calculation
of the Higgs self-energies. Different algorithms for expressing all
two-loop self-energy integrals with arbitrary external momentum in
terms of a minimal set of Master Integrals (MIs) were developed in
refs.~\cite{Weiglein:1993hd,GhinculovYao,Tarasov:1997kx}.  However,
explicit analytical formulae for the MIs can be derived only for
special values of the masses of the particles circulating in the
loops, whereas in the general case a numerical calculation becomes
unavoidable.  A method to compute all the MIs of
ref.~\cite{Tarasov:1997kx} by numerically solving a system of
differential equations in the external momentum was developed in
ref.~\cite{martin_method}, extending earlier results of
ref.~\cite{remiddi}.  A library of routines for the computation of the
MIs with the method of ref.~\cite{martin_method} was then made
available in the package {\tt TSIL}~\cite{TSIL}.

A calculation of the two-loop contributions to the Higgs self-energies
involving the strong gauge coupling or the third-family Yukawa
couplings, based on the methods of refs.~\cite{martin_method,TSIL},
was presented in ref.~\cite{martin_mom}. That calculation goes beyond
the two-loop results implemented in public
codes~\cite{Sven,dsz,bdsz,bdsz2,heidi,dds} in that it includes
external-momentum effects, as well as contributions involving the
D-term-induced EW interactions between Higgs bosons and sfermions.
When combined with the effective-potential results of
ref.~\cite{martin_effpot}, the results of ref.~\cite{martin_mom}
provide an almost-complete two-loop calculation of the Higgs masses in
the MSSM -- the only missing contributions being external-momentum
effects that involve only the EW gauge couplings. However, no code for
the calculation of the MSSM mass spectrum implementing the results of
refs.~\cite{martin_effpot,martin_mom} was ever made available, and the
way those results are organized does not lend itself to a
straightforward implementation in the existing public codes. On one
hand, the $\drbar$ renormalization scheme adopted in
refs.~\cite{martin_effpot,martin_mom} for the parameters of the MSSM
lagrangian does not match the ``mixed on-shell (OS)--$\drbar$'' scheme
adopted in {\tt FeynHiggs}. On the other hand, implementation of the
results of refs.~\cite{martin_effpot,martin_mom} in {\tt SoftSUSY},
{\tt SuSpect} and {\tt SPheno}, which also adopt the $\drbar$ scheme,
is complicated by the fact that in
refs.~\cite{martin_effpot,martin_mom} the running masses of the Higgs
bosons entering the loop corrections are defined by the second
derivatives of the tree-level potential. While this choice amounts to
a legitimate reshuffling of terms between different perturbative
orders, it restricts the applicability of the calculation to rather
specific ranges of renormalization scale where none of the running
Higgs masses -- as defined in refs.~\cite{martin_effpot,martin_mom} --
is tachyonic. Perhaps as a consequence of these complications, a full
decade after the publication of ref.~\cite{martin_mom} its results
have yet to be included in phenomenological analyses of the MSSM Higgs
sector.

In this paper we present a new calculation of the momentum-dependent
part of the two-loop corrections to the neutral Higgs masses of
$\oatas$, i.e.~those involving both the top Yukawa coupling and the
strong gauge coupling. We also compute ``mixed'' two-loop corrections
that we denote by $\oaas$: they involve both the strong gauge coupling
and the EW gauge couplings, under the approximation that the only
non-vanishing Yukawa coupling is the top one. It is natural to
consider these two classes of corrections together, because in both
of them the dominant terms affecting the lightest-scalar mass are
expected to be suppressed by a factor of ${\cal O}(\mz^2/\mt^2)$ with
respect to the zero-momentum $\oatas$ corrections (in practice, we
find that both classes of corrections are considerably more suppressed
than that, but still comparable to each other in size).

In our calculation we rely on the integration-by-parts (IBP) technique
of ref.~\cite{ibp} to express the momentum-dependent loop integrals in
terms of the MIs of ref.~\cite{Tarasov:1997kx}, which we evaluate by
means of the package {\tt TSIL}. We obtain results for both the
$\drbar$ and OS--$\drbar$ renormalization schemes, organized in such a
way that they can be directly implemented in the existing codes for
the computation of the MSSM mass spectrum. We verify that our results
are in full agreement with the ones of ref.~\cite{martin_mom} where
they overlap. After describing our calculation in some detail, we
briefly discuss the numerical impact of the momentum-dependent
$\oatas$ and $\oaas$ corrections to the Higgs masses in a set of
representative points in the MSSM parameter space.

While our paper was in preparation, an independent calculation of the
momentum-dependent $\oatas$ corrections to the neutral Higgs masses in
the MSSM appeared~\cite{loro}, relying on the results of
ref.~\cite{Weiglein:1993hd} for the decomposition of two-loop
integrals into MIs and on the package {\tt SecDec}~\cite{secdec} for
the numerical evaluation of the latter. The results of that
calculation are expressed in the OS--$\drbar$ scheme, and they have
been implemented in the latest version of {\tt FeynHiggs}. Although we
have verified that our results for the contributions of genuine
two-loop diagrams involving the strong-gauge and top-Yukawa couplings
agree numerically with those of ref.~\cite{loro}, we do not reproduce
the overall values of the momentum-dependent $\oatas$ corrections to
the Higgs masses. We trace the reason for the discrepancy to an
inconsistency in ref.~\cite{loro} concerning the definitions of the
wave-function-renormalization (WFR) constants for the Higgs fields and
of the parameter $\tb$.

\section{Neutral Higgs boson masses in the MSSM}
\label{sec:general}

We outline here our calculation of the two-loop corrections to the
masses of the neutral Higgs bosons in the MSSM with real parameters
(we do not consider the possibility of CP violation in the Higgs
sector).
We decompose the neutral components of the two Higgs doublets into
their VEVs plus their CP-even and CP-odd fluctuations as follows
\be 
\label{expansion}
H_1^0 ~\equiv~ \frac{1}{\sq2}\,(v_1 + S_1 + i \, P_1) \, ,
\;\;\;\;\;\;\;
H_2^0 ~\equiv~ \frac{1}{\sq2}\,(v_2 + S_2 + i \, P_2) \, .
\ee
The CP-odd components $P_1$ and $P_2$ combine into the pseudoscalar
$A$ and the neutral would-be Goldstone boson $G^0$. The CP-even
components $S_1$ and $S_2$ combine into the scalars $h$ and $H$. The
squared physical masses of the latter are the two solutions for $p^2$
of the equation
\be 
{\rm det} \left[ \Gamma_S(p^2)
\right] = 0~,
\label{eq:det}
\ee
where $\Gamma_S(p^2)$ denotes the $2\!\times\!2$ inverse-propagator
matrix in the $(S_1,S_2)$ basis, $p$ being the external momentum
flowing into the scalar self-energies. We can decompose
$\Gamma_S(p^2)$ as
\be 
\label{eq:gammaS}
\Gamma_S(p^2) = p^2 ~-~ {\cal M}_0^2 ~-~ \Delta {\cal M}^2(p^2)~,
\ee
where ${\cal M}_0^2$ denotes the tree-level mass matrix written in
terms of renormalized parameters, and $\Delta {\cal M}^2(p^2)$
collectively denotes the radiative corrections. At each order in the
perturbative expansion, the latter include both the contributions of
one-particle-irreducible (1PI) diagrams and non-1PI counterterm
contributions arising from the renormalization of parameters that
enter the lower-order parts of $\Gamma_S(p^2)$. We express ${\cal
  M}_0^2$ in terms of the parameter $\tan\beta$ renormalized in the
$\drbar$ scheme, and of the physical masses of the pseudoscalar and of
the $Z$ boson
\be
{\cal M}_0^2 ~=~ 
\left(\begin{array}{cc}
\cbeq \,\mz^2  + \sbeq\,\ma^2 
& - \sbe\, \cbe \left(\mz^2 + \ma^2 \right) \\ 
- \sbe\, \cbe \left(\mz^2 + \ma^2\right) 
&  \sbeq\, \mz^2 + \cbeq\, \ma^2 
\end{array}\right) 
\label{mzero} \,,
\ee
using (here and thereafter) the shortcuts $c_\phi \equiv \cos\phi$ and
$s_\phi \equiv \sin\phi$ for a generic angle $\phi$.  Neglecting terms
that do not contribute at $\oatas$ or $\oaas$, our choices for the
parameters entering ${\cal M}_0^2$ lead to the following expressions
for the two-loop parts of the individual entries of $\Delta {\cal
  M}^2(p^2)$
\bea
\label{eq:dm11}
\left[\Delta {\cal M}^2(p^2)\right]^{(2)}_{11} &=&
\sbe^2\, {\rm Re}\, \Pi_{\smalla\smalla}^{(2)}(\ma^2) 
+ \cbe^2\,{\rm Re}\, \Pi_{\smallz\smallz}^{(2)}(\mz^2) 
-\Pi_{11}^{(2)}(p^2) 
-\delta \mathcal{Z}_1^{(2)} \left(p^2- \cbeq\, \mz^2 - \sbeq\,\ma^2\right) 
\nonumber\\
&& + \left(1- \sbe^4\right) \frac{~\,T_1^{(2)}}{v_1} 
- \sbe^2 \cbe^2 \frac{~\,T_2^{(2)}}{v_2} 
-2\,\sbe^2\cbe^2\, \left(\mz^2-\ma^2\right)\frac{~\delta \tb^{(2)}}{\tb} 
\,,\\[5pt]
\label{eq:dm12}
\left[\Delta {\cal M}^2(p^2)\right]^{(2)}_{12} &=&
- \sbe\cbe \left[{\rm Re}\,  \Pi_{\smalla\smalla}^{(2)}(\ma^2) 
+ {\rm Re}\, \Pi_{\smallz\smallz}^{(2)}(\mz^2) 
- \sbe^2 \frac{~\,T_1^{(2)}}{v_1} - \cbe^2 \frac{~\,T_2^{(2)}}{v_2} \right] 
-\Pi_{12}^{(2)}(p^2) 
\nonumber \\
&& -\frac{1}{2}\sbe\cbe \left(\mz^2 + \ma^2\right)  
\left[  2\, (\cbe^2-\sbe^2)\frac{~\delta \tb^{(2)}}{\tb} + \delta
\mathcal{Z}_1^{(2)}+\delta\mathcal{Z}_2^{(2)}\right]\,, \\[5pt]
\label{eq:dm22}
\left[\Delta {\cal M}^2(p^2)\right]^{(2)}_{22} &=&
\cbe^2\, {\rm Re}\, \Pi_{\smalla\smalla}^{(2)}(\ma^2) 
+ \sbe^2 \, {\rm Re}\, \Pi_{\smallz\smallz}^{(2)}(\mz^2) 
-\Pi_{22}^{(2)}(p^2) 
-\delta \mathcal{Z}_2^{(2)} \left(p^2- \sbeq\,\mz^2 - \cbeq\, \ma^2 \right)
\nonumber \\
&& - \sbe^2 \cbe^2 \frac{~\,T_1^{(2)}}{v_1} + \left(1- \cbe^4\right)
\frac{~\,T_2^{(2)}}{v_2} +2\,\sbe^2\cbe^2\, \left(\mz^2-\ma^2\right)
\frac{~\delta \tb^{(2)}}{\tb}\,.
\eea
In the equations above, $T^{(2)}_i$ and $\Pi^{(2)}_{ij}$ (with
$i,j=1,2$) denote the two-loop parts of tadpoles and self-energies,
respectively, for the scalars $S_i$, while
$\Pi^{(2)}_{\smalla\smalla}$ and $\Pi^{(2)}_{\smallz\smallz}$ denote
the two-loop parts of the pseudoscalar and $Z$-boson self-energies. In
addition, $\delta {\cal Z}^{(2)}_i$ (with $i=1,2$) in
eqs.~(\ref{eq:dm11})--(\ref{eq:dm22}) denote the two-loop parts of the
WFR counterterms for the Higgs fields $H^0_i$, which we renormalize as
follows:
\be
\label{eq:wfr}
H^0_i ~\longrightarrow~\sqrt{{\cal Z}_i}\,H^0_i ~\simeq~ 
\left(1 + \frac12\,\delta {\cal Z}^{(1)}_i +  \frac12\,\delta {\cal Z}^{(2)}_i
\right)\,H^0_i~,
\ee
where in the expansion of the square root we have again neglected
terms that do not contribute at $\oatas$ or $\oaas$. We adopt a
$\drbar$ definition for the ${\cal Z}_i$, which can then be determined
from the anomalous dimensions of the Higgs fields and from the $\beta$
functions of the couplings entering the anomalous dimensions. Taking
from the general formulae of ref.~\cite{stock} only the terms relevant
to our approximation, we get
\be
\label{eq:dZ}
\delta {\cal Z}^{(1)}_1 ~=~ \delta {\cal Z}^{(2)}_1~=~ 0\,,~~~~~~~
\delta{\cal Z}^{(1)}_2 \,=\, 
- \frac{\at}{4\pi}\,N_c\,\cdot\,\frac1\epsilon\,,~~~~~~~
\delta{\cal Z}^{(2)}_2 \,=\,\frac{\at\as}{(4\pi)^2}\,2\,N_c\,C_F\,\cdot\,
\left(\frac{1}{\epsilon^2}-\frac1\epsilon\right)~,
\ee
where $N_c=3$ and $C_F=4/3$ are color factors, $\epsilon=(4-d)/2\,$ in
dimensional reduction, and the coupling $\at$ entering the one-loop
counterterm $\delta{\cal Z}^{(1)}_2$ is in turn renormalized in the
$\drbar$ scheme. Finally, $\delta\tb^{(2)}$ in
eqs.~(\ref{eq:dm11})--(\ref{eq:dm22}) denotes the two-loop part of the
counterterm for the parameter $\tb$. The choice of a $\drbar$
definition for $\tb$ implies that, in our approximation, its
counterterm can be expressed via the WFR counterterms:
\be
\label{eq:dtanB}
~~~~~~~~~~~~\frac{~\delta\tb^{(\ell)}}{\tb} 
~=~\frac{1}{2}\,\left(\delta{\cal Z}^{(\ell)}_2
-\delta{\cal Z}^{(\ell)}_1\right)~~~~~~~~~~~~(\ell=1,2)~.
\ee

All tadpoles and self-energies in
eqs.~(\ref{eq:dm11})--(\ref{eq:dm22}) include both 1PI two-loop
contributions and non-1PI contributions arising from the
renormalization of the parameters entering their one-loop
counterparts. Since we are focusing on the $\oatas$ and $\oaas$
corrections to the Higgs masses, we need to introduce counterterms
only for the parameters that are subject to $\oas$ corrections, namely
the top mass $m_t$, the stop masses $m_{\stu}$ and $m_{\std}$, the
stop mixing angle $\theta_t$, the soft supersymmetry-breaking
Higgs-stop coupling $A_t$ and the masses $m_{\tilde q_i}$ of all
squarks other than the stops. The latter enter the one-loop tadpoles
and self-energies of the Higgs bosons via D-term-induced EW couplings,
and the one-loop self-energy of the $Z$ boson via the gauge
interaction. In our calculation we neglect all Yukawa couplings (and
hence quark masses) other than the top one,\footnote{The corrections
  to the Higgs masses involving the bottom Yukawa coupling could
  become relevant for large values of $\tb$. When all parameters in
  the bottom/sbottom sector are renormalized in the $\drbar$ scheme,
  the $\oabas$ corrections can be obtained from the corresponding
  results for the $\oatas$ corrections via trivial replacements. On
  the other hand, an OS renormalization of the bottom/sbottom
  parameters would entail additional complications, as discussed in
  refs.~\cite{bdsz2,heidi,dds}. Anyway, the regions of the MSSM
  parameter space where the $\oabas$ corrections to the Higgs masses
  are most relevant are being severely constrained by direct searches
  of Higgs bosons decaying to tau leptons at the LHC~\cite{higgstau}.}
therefore none of the other squarks mix.  We obtained results for the
$\oatas$ and $\oaas$ contributions to tadpoles and self-energies
assuming that the relevant quark/squark parameters are renormalized
either in the $\drbar$ or in the OS scheme. Formulae for the
$\drbar$--OS shift of the parameters in the top/stop sector can be
found, e.g., in appendix B of ref.~\cite{dsz}, while the shifts for
the remaining squark masses can be obtained by setting
$\mt=\theta_t=0$ in the corresponding formulae for the stop masses.
We remark that the right-hand sides of
eqs.~(\ref{eq:dm11})--(\ref{eq:dm22}) are constructed to give finite
entries in the inverse-propagator matrix of the scalars.  Indeed we
have explicitly verified that -- after summing all 1PI and counterterm
contributions -- the $1/\epsilon^2$ and $1/\epsilon$ poles in the
right-hand sides of eqs.~(\ref{eq:dm11})--(\ref{eq:dm22}) cancel out.

In principle, the two-loop contributions to the Higgs inverse
propagator given in eqs.~(\ref{eq:dm11})--(\ref{eq:dm22}) must be
combined with a full calculation of the corresponding one-loop
contributions, and used to determine the physical Higgs masses by
solving directly eq.~(\ref{eq:det}). However, as will be discussed in
the next section, the computing times required for the evaluation of
momentum-dependent two-loop integrals are not negligible. A numerical
search for the solutions of eq.~(\ref{eq:det}) could significantly
slow down the codes for the calculation of the Higgs masses, making
them unsuitable for extensive phenomenological analyses of the MSSM
parameter space. It is therefore convenient to compute the Higgs
masses in two steps, with a procedure similar to the one discussed in
refs.~\cite{brignole,bdsz}. We first call {\tt FeynHiggs}, which
solves eq.~(\ref{eq:det}) including in $\Delta {\cal M}^2(p^2)$ the
full one-loop corrections plus the dominant two-loop corrections of
$\oatas$, $\oabas$ and ${\cal O}(\at^2+\at\ab+\ab^2)$ computed in the
approximation of vanishing external momentum. From {\tt FeynHiggs} we
obtain the scalar masses $\mhb^2$ and $\mHb^2$, and an effective
mixing angle $\alb$ which diagonalizes the loop-corrected scalar mass
matrix at vanishing external momentum. Our full results for the scalar
masses are then obtained by adding to the results of {\tt FeynHiggs}
the momentum-dependent parts of the $\oatas$ corrections and the whole
$\oaas$ corrections:
\be 
m^2_{h,\smallH} ~=~ \overline{m}^2_{h,\smallH} 
~+~ (\Delta m^2_{h,\smallH})^{\at\as,\,p^2}
~+~ (\Delta m^2_{h,\smallH})^{\alpha\as}~.
\label{eq:corrections}
\ee
Concerning the former, we have
\bea
 (\Delta m^2_{h})^{\at\as,\,p^2} &=& 
  c^2_{\beta-\alb}\,\Delta \Pi_{\smalla\smalla}^{(2)}(\ma^2)
\,-\, s^2_{\alb}\, \Delta \Pi_{11}^{(2)}(\mhb^2) 
\,+\, s_{2\alb}\,\Delta \Pi_{12}^{(2)}(\mhb^2)
\,-\, c^2_{\alb}\,\Delta \Pi_{22}^{(2)}(\mhb^2)~,
\label{corrhp2}
\\[5pt]
 (\Delta m^2_{\smallH})^{\at\as,\,p^2} &=& 
  s^2_{\beta-\alb}\,\Delta \Pi_{\smalla\smalla}^{(2)}(\ma^2)
\,-\, c^2_{\alb}\, \Delta \Pi_{11}^{(2)}(\mHb^2) 
\,-\, s_{2\alb}\,\Delta \Pi_{12}^{(2)}(\mHb^2)
\,-\, s^2_{\alb}\,\Delta \Pi_{22}^{(2)}(\mHb^2)~,
\label{corrHp2}
\eea
where we define $\,\Delta\Pi(p^2) \,\equiv\, \Pi(p^2)-\Pi(0)$, and
retain only the real and finite part of the $\oatas$ contributions to
the two-loop self-energies. For what concerns the $\oaas$ corrections,
they contain all terms from eqs.~(\ref{eq:dm11})--(\ref{eq:dm22}):
\bea
 (\Delta m^2_{h})^{\alpha\as} &=& 
s^2_{\alb}\,\left[\Delta {\cal M}^2(\mhb^2)\right]^{\alpha\as}_{11} 
\,-\, s_{2\alb}\,\left[\Delta {\cal M}^2(\mhb^2)\right]^{\alpha\as}_{12} 
\,+\, c^2_{\alb}\,\left[\Delta {\cal M}^2(\mhb^2)\right]^{\alpha\as}_{22}~,
\label{corrhEW}
\\[10pt]
 (\Delta m^2_{\smallH})^{\alpha\as} &=& 
c^2_{\alb}\,\left[\Delta {\cal M}^2(\mHb^2)\right]^{\alpha\as}_{11} 
\,+\, s_{2\alb}\,\left[\Delta {\cal M}^2(\mHb^2)\right]^{\alpha\as}_{12} 
\,+\, s^2_{\alb}\,\left[\Delta {\cal M}^2(\mHb^2)\right]^{\alpha\as}_{22}~,
\label{corrHEW}
\eea
where again we take the real part of all two-loop self-energies.  We
remark that ref.~\cite{loro} proposes an alternative two-step
procedure to include the momentum-dependent parts of the $\oatas$
corrections in {\tt FeynHiggs}, differing from the one outlined
above only by higher-order effects.

Finally, a comment is in order about the dependence of the corrections
to the Higgs masses on the WFR constants.\,\footnote{The
  renormalization of the Higgs fields and of $\tb$ in the calculation
  of the Higgs-mass corrections was also recently discussed, in a
  somewhat different context, in refs.~\cite{haber,sneusven}.} In
principle, the predictions for the physical Higgs masses at a given
order in the perturbative expansion should not depend directly on the
WFR constants (although they could still depend indirectly on them via
the $\tb$ counterterm).  Indeed, if the two-loop contributions to the
inverse-propagator matrix are computed with $p^2$ equal to the
tree-level scalar masses and then rotated to the mass-eigenstate basis
via the tree-level mixing angle, so that the computation is performed
strictly at the two-loop level, the terms in
eqs.~(\ref{eq:dm11})--(\ref{eq:dm22}) that depend explicitly on
$\delta {\cal Z}^{(2)}_i$ drop out of the mass corrections $\Delta
m^2_{h,\smallH}\,$. On the other hand, if the loop-corrected scalar
masses $\mhb^2$ and $\mHb^2$ and the effective mixing angle $\alb$ are
used, as in eqs.~(\ref{corrhp2})--(\ref{corrHEW}) above, or if the
zeroes of the inverse-propagator matrix are determined numerically,
the corrections to the scalar masses retain a dependence on the WFR
counterterms. In our calculation we adopt a $\drbar$ definition for
the WFR, therefore the terms involving $\delta {\cal Z}^{(2)}_i$ are
purely divergent and cancel out against other divergent terms in the
individual entries of the inverse-propagator matrix, hence they do not
show up in eqs.~(\ref{corrhp2}) and (\ref{corrHp2}). If however one
adopts a non-minimal definition of the WFR, Higgs-mass corrections
computed as in eqs.~(\ref{corrhEW}) and (\ref{corrHEW}) will contain
non-vanishing terms that depend on the finite part of $\delta {\cal
  Z}^{(2)}_i$. Albeit formally of higher order in the perturbative
expansion, these terms can be numerically relevant when the
loop-corrected scalar masses differ significantly from their
tree-level values (as is the case for a SM-like scalar $h$ with mass
around 125~GeV).

\section{Calculation of two-loop diagrams with nonzero 
momentum}
\label{sec:calc}

The computation of the two-loop corrections to the neutral MSSM Higgs
masses considered in this paper requires the knowledge of the tadpole
and self-energy diagrams entering
eqs.~(\ref{eq:dm11})--(\ref{eq:dm22}).  While the strategy for the
computation in the zero-momentum approximation is well known, the
evaluation of the self-energies with arbitrary external momentum is
more involved. We illustrate in this section the details of our
calculation, which we performed in a fully automated way.

The relevant diagrams are generated with
\texttt{FeynArts}\cite{feynarts}, using a modified version of the
original MSSM model file~\cite{Hahn:2001rv} that implements the QCD
interactions in the background field gauge. The diagrams contributing
to the vacuum polarization of the $Z$ boson are contracted with a
suitable projector in order to single out their transverse part.  The
color factors are simplified with a private package and the Dirac
algebra is handled by \texttt{FORM}~\cite{form}.
The computation is performed in dimensional reduction, which we can
implement in this case by generating the diagrams in dimensional
regularization and replacing, in each diagram involving an internal
$d$-dimensional gluon, $g^{\mu\nu} \to g^{\mu\nu} +
g^{\hat\mu\hat\nu}$ (where $g^{\hat\mu\hat\nu}$ is the
$2\epsilon$-dimensional metric tensor) in order to include the
corresponding $\epsilon$-scalar contribution.  We are then left with
Feynman integrals of the form
\begin{eqnarray}
  \int d^dk_1 d^dk_2 \frac{(k_1^2)^\alpha (k_2^2)^\beta (k_1\cdot
    p)^\gamma (k_2\cdot p)^\delta (k_1\cdot k_2)^\eta}{D_1^{a_1}
    D_2^{a_2} D_3^{a_3} D_4^{a_4} D_5^{a_5}}\,,
  \label{eq:Mtopology}
\end{eqnarray}
where $\alpha\,, \ldots\,, \eta,\, a_1\,, \ldots\,, a_5$ are positive
(or zero) integer exponents and the $D_i$'s are defined as
\begin{equation}
  D_1 = k_1^2 - m_1^2\,, \enskip D_2 = (k_1-p)^2 - m_2^2\,, \enskip
  D_3 = k_2^2 - m_3^2\,, \enskip D_4 = (k_2-p)^2 - m_4^2\,, \enskip
  D_5 = (k_1-k_2)^2 - m_5^2\,. \nonumber
\end{equation}
Integrals belonging to the class above are in general not linearly
independent of each other.  When the scalar products in the numerator
are expressed in terms of the denominators, powers of a $D_i$ present
in the original integral might cancel against a $D_i$ in the
numerator, possibly generating a Feynman integral in which $D_i$ does
not appear, i.e.\ in which the corresponding line has been shrunk to a
point.  For given $a_i$'s and high enough $\alpha,\ldots,\eta$, some
$D_i$'s may acquire negative exponents.  The computation of a Feynman
integral of the type in eq.~(\ref{eq:Mtopology}) therefore reduces to
the evaluation of a number of integrals of the form
\begin{eqnarray}
  \int \frac{d^dk_1 d^dk_2}{D_1^{n_1} D_2^{n_2} D_3^{n_3} D_4^{n_4} D_5^{n_5}}\,,
  \label{eq:Mtopology2}
\end{eqnarray}
where the exponents $n_i\in \mathbb{Z}$.
In the present case, one has to evaluate $\mathcal{O}(300)$ different
Feynman integrals.

There exists a convenient procedure for dealing with such large
numbers of different Feynman integrals in a more efficient way than
their direct evaluation.
Dimensionally regularized integrals, at arbitrary loop order and with
arbitrary number of external legs, satisfy identities of the IBP
type~\cite{ibp}.  These identities are linear relations that connect
integrals with different sets of exponents $\{n_1,\ldots,n_5\}$.
After a set of independent integrals, the MIs, has been identified,
all the remaining integrals can then be reduced to linear combinations
of the MIs, the coefficients being rational functions of the masses,
the kinematic invariants and the space-time dimension $d$.
One practical advantage of such a procedure is its ``divide and
conquer'' spirit.  On the one hand few MIs encode the analyticity
properties (singularities, thresholds, branch cuts) of the problem
under consideration. On the other hand, the evaluation of the large
number of different Feynman integrals entering a computation is
reduced to a problem of linear algebra, which can easily be automated,
if the MIs are known.
We perform the reduction to MIs with the public code
\texttt{REDUZE}~\cite{reduze}, which implements the Laporta
algorithm~\cite{laporta}, and produces the IBP identities relevant to
our case.

The evaluation of the MIs is in general a complicated problem and can
proceed via different techniques, like the integration in the Feynman,
Schwinger or Mellin-Barnes representations.
A remarkable consequence of the aforementioned IBP relations is that
the MIs obey linear systems of first-order differential equations
(DEs) in the kinematic invariants, which provide an alternative means
for their computation~\cite{DEs}. Finding the analytic solution of the
DEs for arbitrary $d$ is possible only in some simple cases. In more
general cases the MIs are expanded in powers of $\epsilon = (4-d)/2$,
giving rise to a (generally coupled) system of DEs for the expansion
coefficients.
In the limit of vanishing external momentum, two-loop self-energies
become two-loop vacuum diagrams, which reduce via IBP to linear
combinations of only one genuine two-loop MI and
products of the one-loop one-propagator MI~\cite{vacuum2L}. 
Two-loop self-energies with arbitrary external momentum, in the
general case with five different masses in the loops, reduce via IBP
to linear combinations of 30 MIs~\cite{Tarasov:1997kx}.  The finite
part of such MIs can be expressed in terms of four functions, in
addition to the well-known one-loop MIs.\footnote{ In the presence of
  infrared divergences associated to loops of massless quarks, the IBP
  reduction of the considered diagrams to MIs requires two additional
  functions. Their expression in terms of logarithms and
  polylogarithms can be obtained from the results of
  ref.~\cite{magnus}.  }
As already mentioned, analytic solutions for such functions have been
derived only for special patterns of up to two internal masses (only
one function is known in a particular case with three different
masses).  On the other hand, the diagrams that in our approximation
contribute to the self-energies entering
eqs.~(\ref{eq:dm11})--(\ref{eq:dm22}) require the knowledge of MIs
with up to four different masses, the most complicated ones being
those involving simultaneously $\t\,,\tu\,,\td$ and $\g$.

In our computation we rely on the package {\tt TSIL}~\cite{TSIL},
which implements (besides all the analytically known cases) the
numerical solution of the DEs for the two-loop self-energy MIs.  The
method of refs.~\cite{remiddi,martin_method} is based on the fact that
the value at $p^2=0$ (or the expansion for small $p^2$ in the case of
logarithmically divergent integrals) is known for each function and
can be used to build the set of initial conditions needed for the
solution of the DEs.
In the computation of the self-energies entering
eqs.~(\ref{eq:dm11})--(\ref{eq:dm22}) we need to evaluate the
corresponding MIs at $p^2=\mz^2\,,\ma^2\,,\mh^2\,,\mH^2$. Given that
we include the contribution of light quarks to $\Pi_{\smallz\smallz}$
(in the approximation $m_q=0$) and that $\ma$ is a free parameter, 
it is clear that the way thresholds are
handled in the numerical evaluation of the MIs is of particular
relevance.  In the DEs approach, the physical two- and three-particle
thresholds show up, together with the pseudothresholds, as poles in
the coefficients of the DEs.
{\tt TSIL} overcomes the numerical instabilities related to such poles
by displacing the $p^2$-integration contour in the upper half-plane
when the momentum is above the smallest (pseudo)threshold.  The
evaluation at (or very close to) the (pseudo)thresholds is performed
through a variant of the algorithm, which is slightly less efficient
but ensures reliable results in such critical cases.
As an example, the time needed on an Intel Core i7-4650U CPU for the
evaluation of the \emph{complete set} of MIs, for any of the mass
patterns entering the self-energies, ranges between $5 \times
10^{-4}\,s$ and $8 \times 10^{-2}\,s$, the latter being the typical
time for $\sqrt{p^2}$ close or equal to the heavy stop pair threshold
and to the three-particle (pseudo)thresholds $m_{\tilde{t}_i}+\mg \pm
\mt$.
In ref.~\cite{TSIL} the relative accuracy of {\tt TSIL} is claimed to
be better than $10^{-10}$ for generic cases, or worse in cases with
large mass hierarchies.
Being {\tt TSIL} a package dedicated to the evaluation of the MIs for
two-loop self-energy diagrams, it is not surprising that its speed and
accuracy prove much better than those quoted in ref.~\cite{loro},
where the general-purpose package \texttt{SecDec} is used and, in the
most complicated case, $100\,s$ are needed in order to reach a
relative accuracy of at least $10^{-5}$ close to a threshold.

\section{Numerical examples}
\label{sec:num}

\begin{figure}
\begin{center}
\includegraphics[width=.9\textwidth]{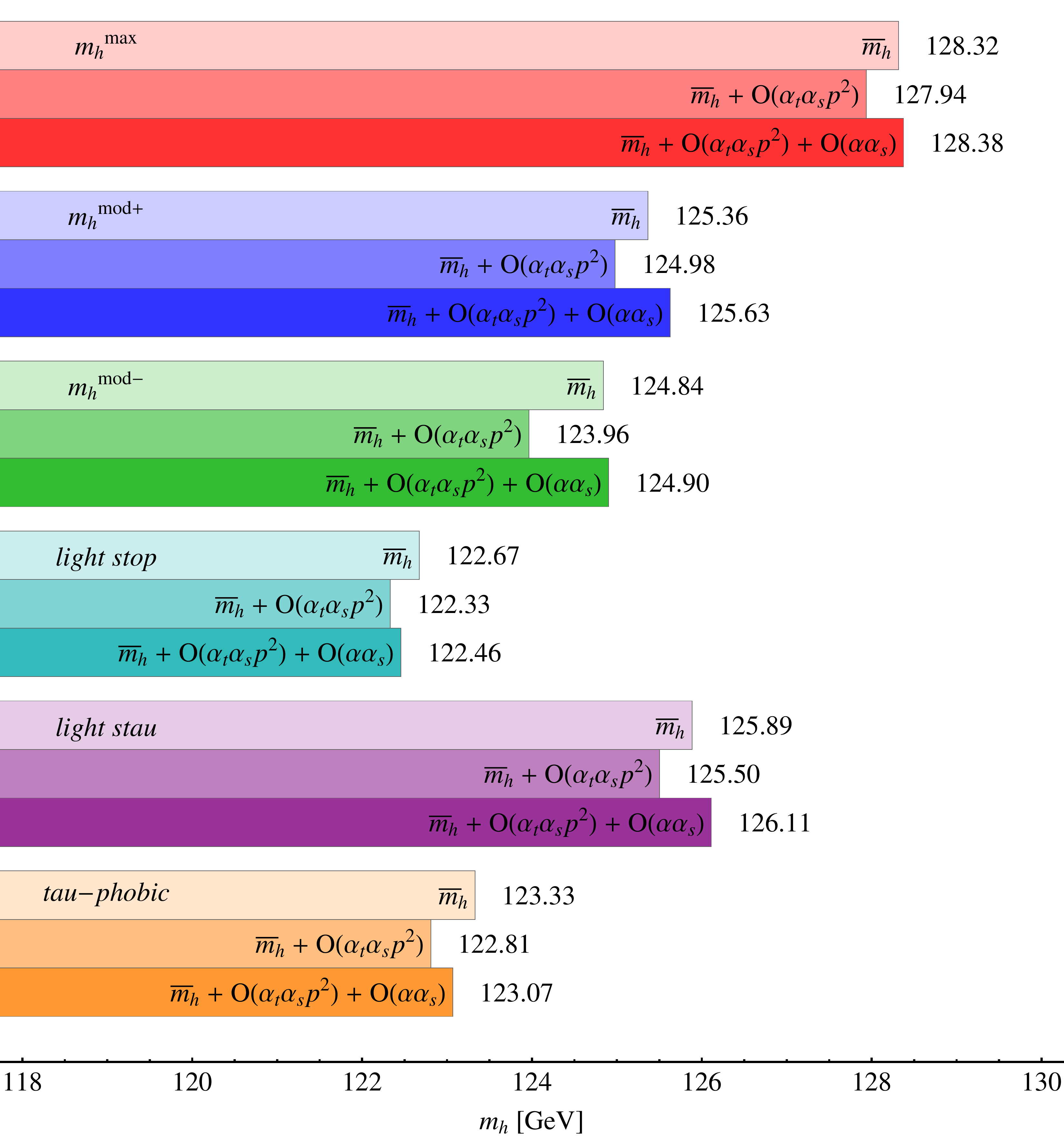}
\caption{Predictions for the mass of the lightest scalar $h$ in the
  six benchmark scenarios of ref.~\cite{Carena:2013qia}, for
  $\ma=500~\gev$ and $\tan\beta=20$.  For each scenario, the three
  bars show: the ``unperturbed'' mass $\overline{m}_h$ computed with
  {\tt FeynHiggs~\fhversion} (upper), the inclusion of the
  momentum-dependent part of the $\oatas$ corrections (middle) and the
  additional inclusion of the whole $\oaas$ corrections (lower).  From
  top to bottom, the considered scenarios are $\mhmax$ (red),
  $\mh^{\rm mod+}$ (blue), $\mh^{\rm mod-}$ (green), {\em light stop}
  (turquoise), {\em light stau} (purple), {\em tau-phobic} (orange).}
  \label{fig:bars}
\end{center}
\end{figure}

In this section we assess the numerical impact of the
momentum-dependent part of the $\oatas$ corrections and of the whole
$\oaas$ corrections on the predictions for the neutral Higgs-boson
masses in the MSSM.
We focus on six benchmark scenarios introduced in
ref.~\cite{Carena:2013qia}, which identify regions in the MSSM
parameter space compatible with the current bounds from SUSY-particle
searches and with the requirement that the predicted value of $\mh$
agrees, within the theoretical uncertainty of $\pm 3~\gev$ estimated
in refs.~\cite{Degrassi:2002fi,Allanach:2004rh}, with the mass of the
SM-like Higgs boson discovered at the LHC.~\footnote{We omit a seventh
  scenario from ref.~\cite{Carena:2013qia}, the so-called {\em
    low-$\mH$} scenario. Also, the parameters in the {\em light-stop}
  scenario are modified as in ref.~\cite{Bagnaschi:2014zla}, to
  account for newer exclusion bounds from direct stop searches at the
  LHC.}

In our numerical examples we adopt the mixed OS--$\drbar$ scheme
described in section~\ref{sec:general}. The SM input parameters are
chosen as the pole top mass $m_t = 173.2~\gev$, the running bottom
mass $m_b(m_b) = 4.2~\gev$, the Fermi constant $G_F = 1.16639 \times
10^{-5}~\gev^{-2}$, the strong gauge coupling $\alpha_s(\mz) = 0.118$,
and the pole gauge-boson masses $\mz = 91.1876~\gev$ and $\mw =
80.385~\gev$.
To compute the scalar masses $\overline{m}^2_{h,\smallH}$ and the
effective mixing angle $\overline{\alpha}$ entering the corrections
in eqs.~(\ref{eq:corrections})--(\ref{corrHEW}), we call {\tt
  FeynHiggs} version {\tt \fhversion}. We use default values for all
settings with the exception of {\tt runningMT\! = \!0}, i.e.~the top
mass in the radiative corrections is identified with the pole mass (to
match the renormalization conditions imposed both in our OS--$\drbar$
calculation and in the one of ref.~\cite{loro}). By default, the
renormalization scale associated to the $\drbar$ definition of the
Higgs WFR and of $\tb$ is fixed as $\muR=m_t$.

In figure~\ref{fig:bars} we present our predictions for the
lightest-scalar mass $m_h$ in the six benchmark scenarios. We choose
$\ma = 500~\gev$ and $\tan\beta=20$, so that the lightest scalar $h$
is SM-like, the bound on its tree-level mass is saturated, and the
corrections controlled by the bottom Yukawa coupling, which we do not
compute beyond the approximations of {\tt FeynHiggs}, are not expected
to be particularly relevant.
For each scenario we show three bars: the upper one represents the
``unperturbed'' mass $\overline{m}_{h}$, obtained from {\tt FeynHiggs}
without additional corrections; the middle bar includes the effect of
the momentum-dependent part of the $\oatas$ corrections, i.e.~the
$(\Delta\mh^2)^{\at\as,\,p^2}\!$ defined in eq.~(\ref{corrhp2});
finally, the lower bar represents our final result for $m_h$, and
includes the effects of both the momentum-dependent part of the
$\oatas$ corrections and the $\oaas$ corrections, i.e.~the $(\Delta
\mh^2)^{\alpha\as}$ defined in eq.~(\ref{corrhEW}).

Figure~\ref{fig:bars} shows that, in all considered scenarios, the
momentum-dependent part of the $\oatas$ corrections and the whole
$\oaas$ corrections can shift the prediction for $m_h$ by several
hundred MeV each (the largest shifts, of about $\pm 1$~GeV, occur in
the $m_h^{\rm mod-}$ scenario).  However, in all of our examples the
two classes of corrections happen to be similar to each other in
magnitude, and to enter the prediction for $m_h$ with opposite
signs. As a result, their combined effect is always fairly small, less
than $\pm 300$~MeV.

\begin{figure}[p]
$$\includegraphics[width=0.47\textwidth]{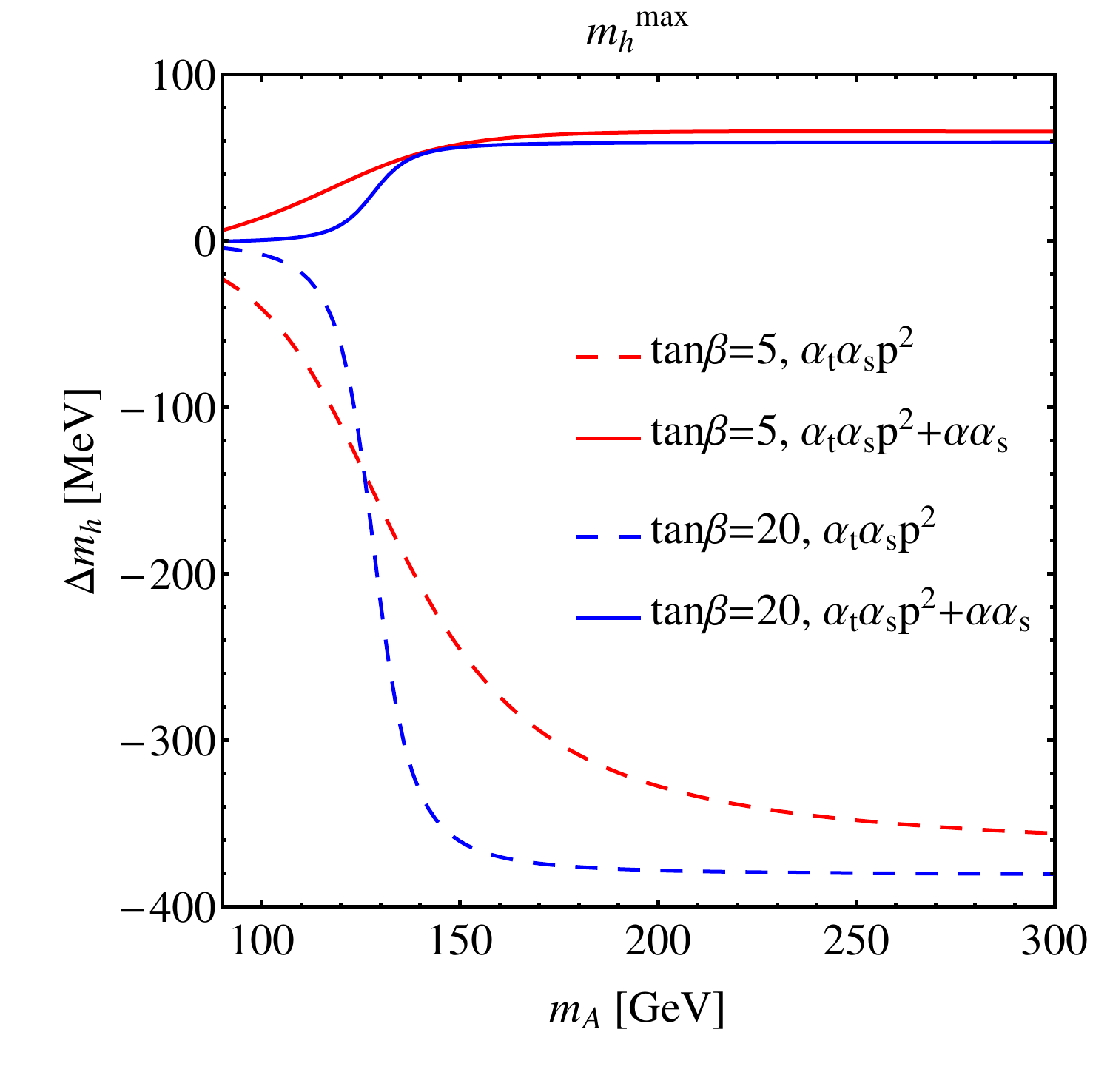}\qquad
\includegraphics[width=0.47\textwidth]{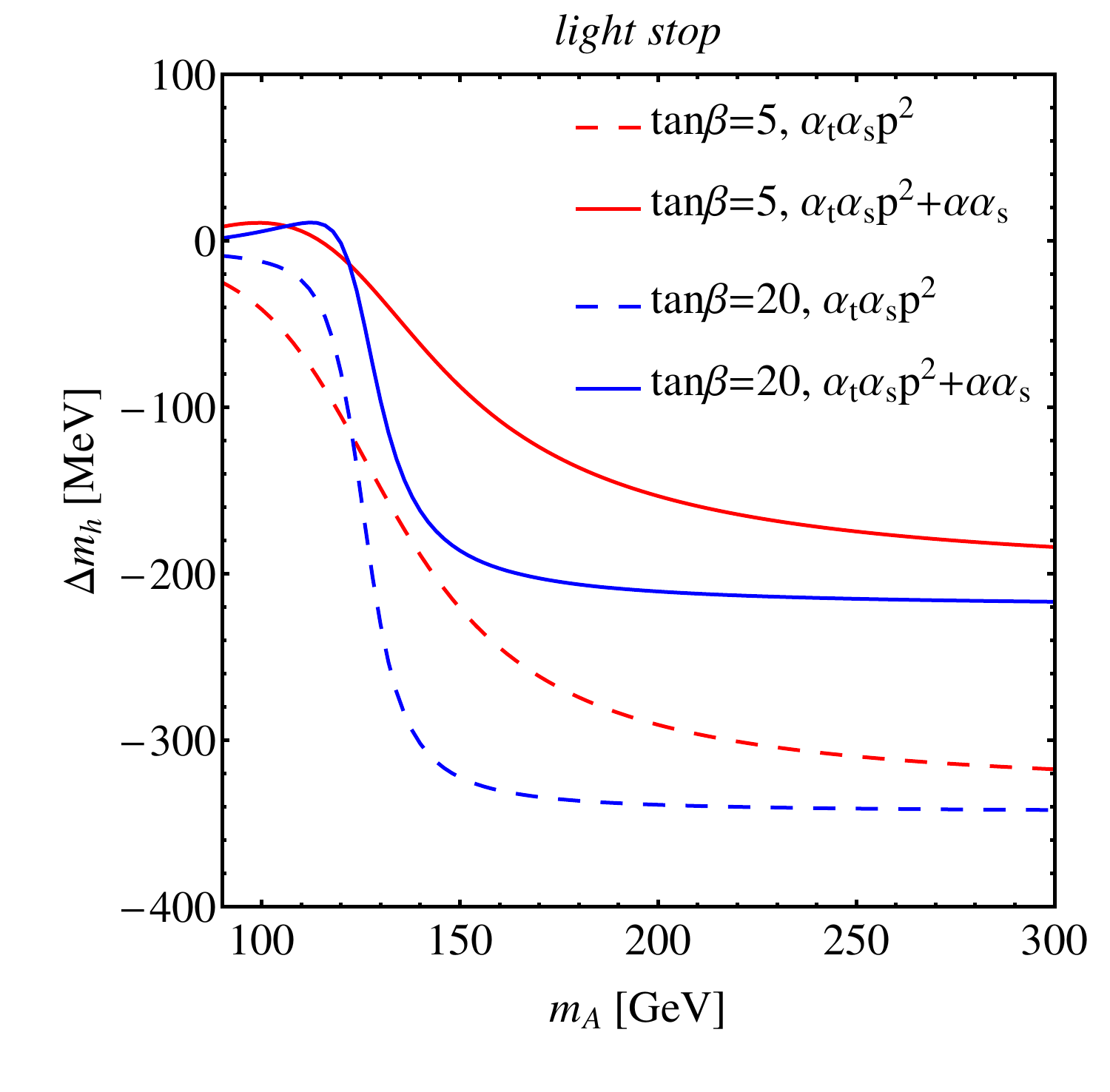}$$
\vspace*{-1cm}
\caption{Corrections to the lightest-scalar mass as function of $\ma$,
  for $\tan\beta=5$ (red) and for $\tan\beta=20$ (blue). The other
  MSSM parameters are chosen as in the $m_h^{\rm max}$ scenario (left)
  or as in the {\em light-stop} scenario (right).  The dashed lines
  represent the contribution of the sole momentum-dependent part of
  the $\oatas$ corrections, the solid lines include both the
  momentum-dependent $\oatas$ corrections and the $\oaas$
  corrections.}
  \label{fig:dmhvsma}
\end{figure}

\begin{figure}[p]
$$\includegraphics[width=0.47\textwidth]{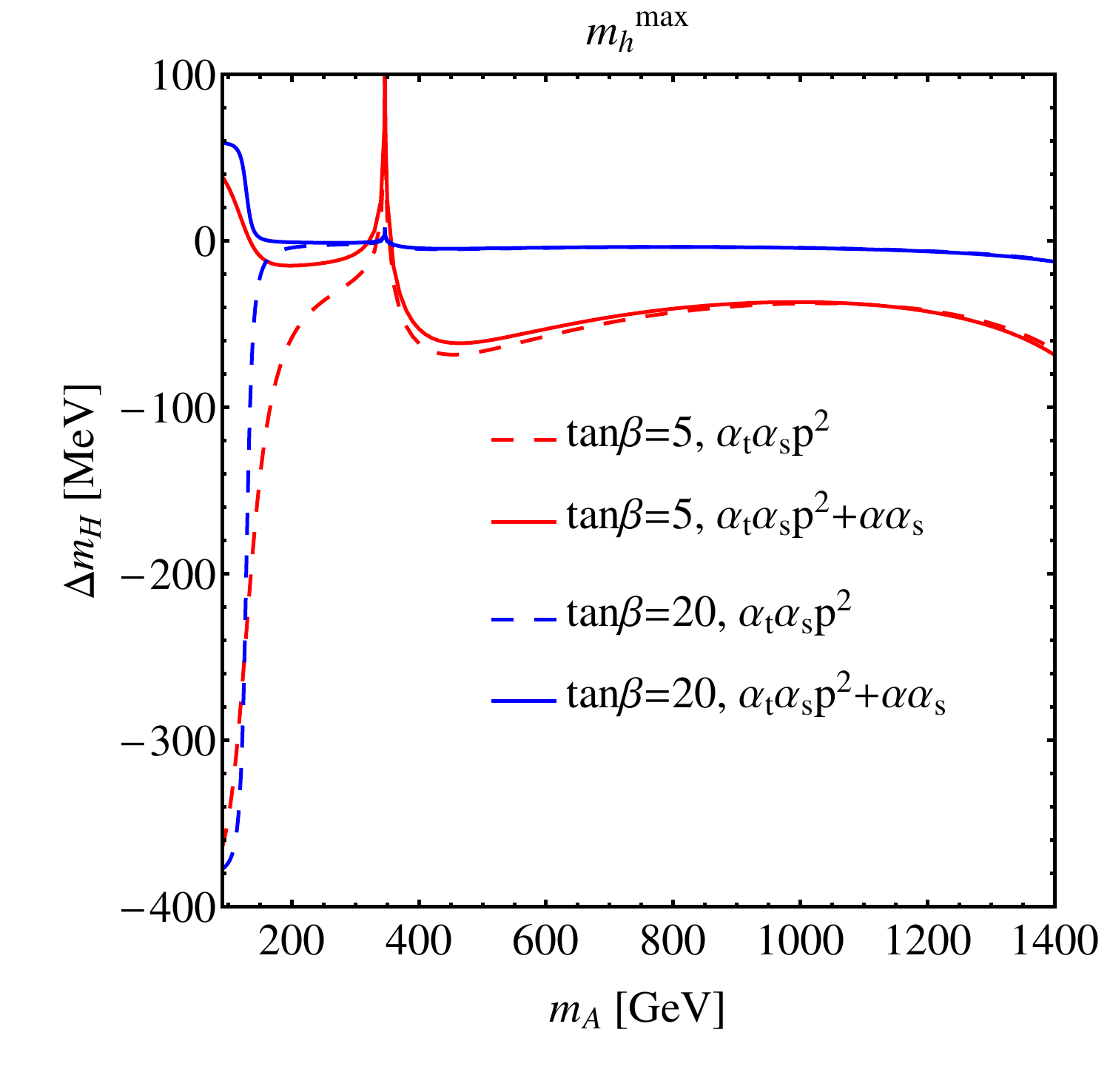}\qquad
\includegraphics[width=0.47\textwidth]{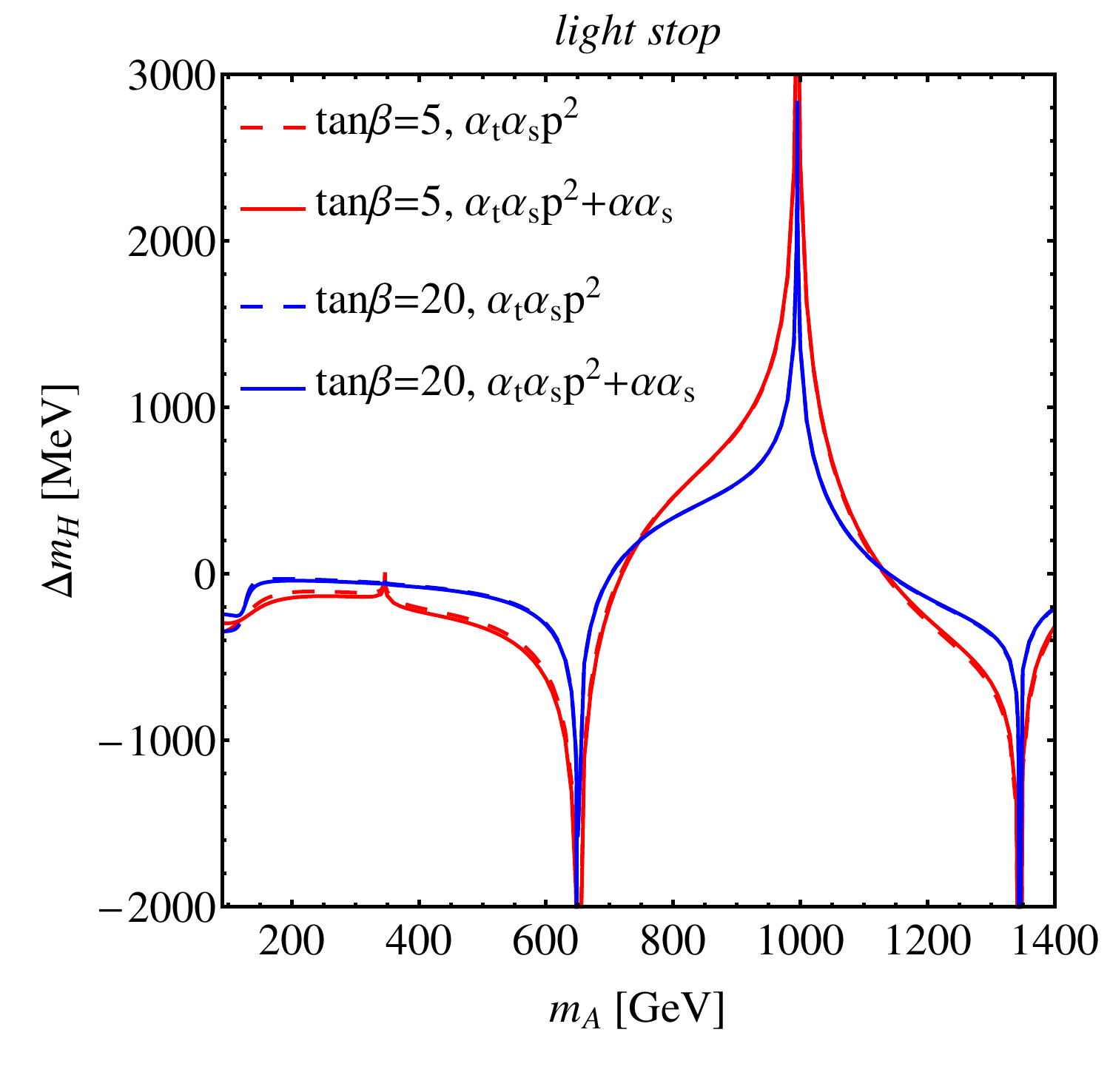}$$
\vspace*{-1cm}
\caption{Same as figure~\ref{fig:dmhvsma} for the corrections to the 
    heaviest-scalar mass.}
  \label{fig:dmhhvsma}
\end{figure}

In figure~\ref{fig:dmhvsma} we illustrate the impact of the
momentum-dependent $\oatas$ corrections and of the $\oaas$ corrections
on the prediction for the lightest-scalar mass as a function of $\ma$,
and in figure~\ref{fig:dmhhvsma} we do the same for the
heaviest-scalar mass.  In each figure, the MSSM parameters for the
left plot are chosen as in the $m_h^{\rm max}$ benchmark scenario of
ref.~\cite{Carena:2013qia}, while for the right plot they are chosen
as in the modified {\em light-stop} scenario. In each plot, the dashed
lines represent the contribution of the sole momentum-dependent part
of the $\oatas$ corrections, while the solid lines include both the
momentum-dependent $\oatas$ corrections and the $\oaas$
corrections. The red lines were obtained with $\tb=5$ while the blue
lines were obtained with $\tb=20$.

Figure~\ref{fig:dmhvsma} shows that the corrections to the
lightest-scalar mass are negligible at low values of $\ma$, but they
become larger and essentially independent of $\ma$ as the latter
increases. The transition to this ``decoupling'' regime -- where the
lightest scalar has SM-like couplings and its mass is insensitive to
the value of $\ma$ -- is sharper for larger values of $\tb$. In both
the $m_h^{\rm max}$ and {\em light-stop} scenarios, the
momentum-dependent $\oatas$ effects decrease $\mh$ by
$300\!-\!400$~MeV at large $\ma$. However, as already seen in
figure~\ref{fig:bars}, the $\oaas$ effects enter the prediction for
$\mh$ with comparable magnitude but opposite sign, significantly
reducing the total size of the correction.

Figure~\ref{fig:dmhhvsma} shows that for low values of $\ma$, where
the heaviest scalar is the one with SM-like couplings, the corrections
to its mass are comparable to the ones that affect the lightest-scalar
mass in the decoupling region. On the other hand, for larger values of
the pseudoscalar mass -- where $\mH \approx \ma$ -- the corrections to
the heaviest-scalar mass show a series of spikes, related to the
opening of real-particle thresholds in diagrams that involve a virtual
gluon. The first spike is visible in correspondence with $\ma =
2\,\mt$ in the plot on the left for the $m_h^{\rm max}$
scenario. More-pronounced spikes (note the different scale on the $y$
axis) are visible in correspondence with $\mH = 2\,m_{\tilde{t}_1}$,
$\mH = m_{\tilde{t}_1}\!+m_{\tilde{t}_2}\,$ and $\mH =
2\,m_{\tilde{t}_2}\,$ in the plot on the right for the {\em
  light-stop} scenario. Analogous spikes would appear at larger values
of $\ma$ in the $m_h^{\rm max}$ scenario, where the stops are
heavier. We stress that our results are not reliable in the vicinity
of these thresholds: the two-loop correction to the heaviest-scalar
mass is actually divergent there, and the height of the spikes in the
plots carries no physical meaning. A more sophisticated analysis,
taking into account the widths of the virtual particles in the loops
as well as non-perturbative QCD effects, would be necessary around the
thresholds, but it is beyond the scope of our calculation.

Figure~\ref{fig:dmhhvsma} also shows that, in the decoupling region
and away from the thresholds, the corrections to the heaviest-scalar
mass amount at most to a few hundred MeV, and they decrease in size
with increasing $\tb$. Moreover, the effect of the $\oaas$ corrections
is negligible (the dashed and solid lines are practically overlapping
in the plots). Inspection of our analytic formulae shows that, in the
decoupling limit, the $\oaas$ corrections to $\mH$ are suppressed by
one or two powers of $\tb$, whereas the $\oatas$ corrections contain
unsuppressed contributions proportional to the square of the
superpotential Higgs-mass parameter $\mu$. While corrections of this
size might be considered negligible in comparison with the value of
$\mH$ itself, they are not entirely irrelevant when compared to the
difference $\mH-\ma$, which can be of the order of a few GeV and is
the quantity of interest when a large physical mass for the
pseudoscalar is taken as input in the calculation.

\section{Comparison with earlier calculations}
\label{sec:comp}
The way we compute the $\oatas$ and $\oaas$ corrections to the entries
of the inverse-propagator matrix for the neutral scalars allows for a
relatively easy comparison with earlier calculations. We first
renormalize all the relevant parameters in the $\drbar$ scheme,
i.e.~we introduce minimal counterterms that, by definition, subtract
only powers of $1/\epsilon\,,$ multiplied by coefficients that should be
polynomial in the $\drbar$-renormalized masses and couplings. In a
second step, we convert our results to the mixed OS--$\drbar$ scheme
adopted in {\tt FeynHiggs}, replacing the $\drbar$ top/stop parameters
entering the one-loop part of the corrections with the corresponding
OS parameters plus the finite one-loop shifts given in
ref.~\cite{dsz}.

As a first obvious check, we took the limit of vanishing external
momentum in the scalar self-energies entering the $\oatas$ corrections
and we compared our results with those in ref.~\cite{dsz}, finding
full agreement. We also successfully compared the $\oaas$ corrections
at vanishing external momentum with the results of an independent
calculation based on the effective-potential techniques of
ref.~\cite{dsz}. Note, however, that this comparison does not cover
the $\oaas$ contributions to the $Z$-boson self-energy. Concerning the
latter, we checked that we can reproduce the result of
ref.~\cite{Djouadi:1993ss} for the subset of two-loop diagrams that
involve only quarks and a gluon, taking into account the fact that
ref.~\cite{Djouadi:1993ss} employed dimensional regularization.

We then compared our results for the momentum-dependent corrections
with those of ref.~\cite{martin_mom}, where the two-loop calculation
of the Higgs masses was performed entirely in the $\drbar$ scheme.  As
mentioned in section~\ref{sec:intro}, the Higgs-mass corrections in
ref.~\cite{martin_mom} are organized in a different way with respect
to our calculation, therefore we could compare only at the level of
individual two-loop self-energies for scalars and pseudoscalars (the
two-loop self-energy for the $Z$ boson was not computed in
ref.~\cite{martin_mom}). Rotating our scalar self-energies from the
$(S_1,S_2)$ basis to the $(h,H)$ basis with the tree-level mixing
angle defined in ref.~\cite{martin_mom}, we reproduce perfectly the
results for the ``top/gluon'' and ``top/stop/gluino'' contributions to
$\Delta \Pi_{hh}$ shown in figure~2 of that paper. This provides a
full cross-check of the momentum-dependent $\oatas$ contribution to
the self-energy, as well as a partial check of the $\oaas$
contribution, restricted to diagrams involving the stop squarks (the
diagrams involving the other squarks are included in the ``others''
line in the above-mentioned figure).
We also checked the analogous contributions to $\Delta
\Pi_{h\smallH}$, $\Delta \Pi_{\smallH\smallH}$ and $\Delta
\Pi_{\smalla\smalla}$ against results provided by the author of
ref.~\cite{martin_mom}, finding again perfect numerical
agreement. Although our calculation and the one in
ref.~\cite{martin_mom} both use {\tt TSIL} to compute the MIs, and
thus cannot be considered entirely independent, the agreement in the
results for the self-energies gives us confidence that the computation
of two-loop Feynman diagrams in terms of MIs and the $\drbar$
subtraction of their divergences are correct in both papers.

Our results for the momentum-dependent $\oatas$ corrections in the
mixed OS--$\drbar$ scheme can in turn be compared with those of
ref.~\cite{loro}. To start with, we compared our two-loop 1PI
contributions to the scalar and pseudoscalar self-energies with
analogous results provided by the authors of ref.~\cite{loro}, and we
found agreement within the accuracy of the sector-decomposition
procedure used to compute the loop integrals in that paper. The
successful comparison between two sets of self-energies in which the
loop integrals were evaluated with {\tt TSIL} and {\tt SecDec},
respectively, validates the results for the two-loop MIs, thus
reinforcing our cross-check of ref.~\cite{martin_mom}. On the other
hand, our results for the momentum-dependent $\oatas$ corrections to
the Higgs masses, obtained by combining the 1PI diagrams with all the
necessary counterterm contributions, differ significantly from the
ones in ref.~\cite{loro}. Considering for example the $\mhmax$
scenario discussed in the previous section, we find that for large
$\ma$ the lightest-scalar mass is subject to a negative correction of
about $350\!-\!400$~MeV (depending on $\tb$, see the left plot in
figure~\ref{fig:dmhvsma}), whereas the corresponding correction in
ref.~\cite{loro} is also negative but quite smaller, about
$50\!-\!60$~MeV (see the upper plot in figure 7 of that paper).  We
traced the reason for this discrepancy to an inconsistency in
ref.~\cite{loro}, related to the renormalization conditions for the
Higgs fields and for $\tb$.

In the $\drbar$ scheme, the WFR counterterm for each field $H^0_i$ can
be related to the divergent part of the derivative of the scalar
self-energy with respect to the external momentum:
\be
\label{eq:dZloro}
~~~~~~~~~~~~\delta{\cal Z}^{(\ell)}_i ~=~ 
- \left[ \frac{d\, {\rm Re}\,\Pi^{(\ell)}_{ii}(p^2)}{dp^2}
\right]^{\rm div}~~~~~~~~~~~~(\ell=1,2)~.
\ee
Indeed, when all parameters entering the one-loop part of the scalar
self-energies are renormalized in the $\drbar$ scheme,
eq.~(\ref{eq:dZloro}) leads to the $\drbar$ WFR counterterms given in
eq.~(\ref{eq:dZ}), in accordance with the anomalous dimensions of the
Higgs fields given in ref.~\cite{stock}.
However, in the mixed OS--$\drbar$ scheme of ref.~\cite{loro} the
top/stop parameters in the one-loop self-energies are renormalized
OS. In that case, the use of eq.~(\ref{eq:dZloro}) to determine the
WFR counterterms leads to
\be
\label{eq:dZwrong}
{\cal Z}_2^{\,\citeloro} ~=~1 
~-~ \frac{~\at^{\smallOS}}{4\pi}\,N_c\,\cdot\,\frac1\epsilon
~+~ \frac{\at\as}{(4\pi)^2}\,2\,N_c\,C_F\,\cdot\,
\left(\frac{1}{\epsilon^2}-\frac1\epsilon\right)
~-~ \frac{\at}{2\pi}\,N_c\,
\frac{\delta m_t}{m_t}\,\cdot\,\frac1\epsilon~,
\ee
where $\at^{\smallOS}$ is a scale-independent coupling extracted from
the pole top mass, and $\delta m_t$ is the finite one-loop shift for
the top mass given in eq.~(B2) of ref.~\cite{dsz}. By converting the
coupling $\at^{\smallOS}$ in the one-loop term of
eq.~(\ref{eq:dZwrong}) into the corresponding $\drbar$ coupling, it is
easy to see that ${\cal Z}_2^{\,\citeloro}$ differs from the $\drbar$
WFR constant in eq.~(\ref{eq:dZ}) by a finite two-loop term:
\be
\label{eq:dZshift}
{\cal Z}_2^{\,\citeloro} ~=~ 
{\cal Z}_2^{\,\smalldr} ~+~ \frac{\at}{2\pi}\,N_c\,
\frac{\delta_\epsilon m_t}{m_t}~,
\ee
where $\delta_\epsilon m_t$ denotes the part proportional to
$\epsilon$ in the top self-energy regularized with dimensional
reduction:
\bea
\label{eq:depsmt}
\frac{\delta_\epsilon m_t}{m_t} &=& 
\frac{\alpha_s}{4 \pi} \,C_F\, \left\{
 -\frac32\,\ln^2 \frac{\mt^2}{\muR^2} +5\,\ln \frac{\mt^2}{\muR^2} 
- \frac{~\pi^2}{4} - 9
- \frac{\g}{\mt^2} \left(  \frac12 \, \ln^2 \frac{\g}{\muR^2} 
- \ln \frac{\g}{\muR^2} +\frac{~\pi^2}{12} +1 \right) \right. \nn \\
&&\nn\\
&&\qquad \qquad +  \,\frac12\left[ ~
\frac{\g +\mt^2 - \tu - 2 \,\sdt \,\mg\, \mt  }{\mt^2} 
\,B_\epsilon(\mt^2,\g,\tu)  \right. \nn \\
&&\nn\\
&& \qquad \qquad \qquad \left.\left.
\,+ \,\frac{\tu}{\mt^2} 
\left( \frac12 \, \ln^2 \frac{\tu}{\muR^2} - \ln \frac{\tu}{\muR^2} 
+\frac{~\pi^2}{12} +1 \right)
\vphantom{ \left( \frac{\g}{\muR^2} \right)}
 +~(\tilde t_1 \rightarrow \tilde t_2,
~~ \sdt\rightarrow-\sdt) \right] \right\}\,.
 \nn \\
\eea
In the equation above $\muR$ is the renormalization scale associated
to the Higgs WFR and to $\tb$, while $B_\epsilon(s,x,y)$ denotes the
coefficient of $\epsilon$ in the expansion of the Passarino-Veltman
function $B_0$. An explicit expression for $B_\epsilon$ can be found,
e.g., in eq.~(2.31) of the {\tt TSIL} manual~\cite{TSIL}.

In the calculation of ref.~\cite{loro}, where the top/stop parameters
entering the one-loop part of the corrections are directly
renormalized OS instead of being first renormalized in the $\drbar$
scheme and then converted to the OS scheme via a finite shift, the
two-loop self-energies and tadpoles contain terms proportional to
$\delta_\epsilon m_t$. Such terms would drop out of the final result
for the renormalized inverse-propagator matrix if
eq.~(\ref{eq:dZshift}) was used to obtain the correct $\drbar$
definition for the WFR constant ${\cal Z}_2^{\,\smalldr}$, and
consequently for $\delta\tb$, but they survive if the WFR constant is
defined as in eq.~(\ref{eq:dZwrong}). To prove that these terms are
indeed at the root of the observed discrepancies, we modified our own
calculation, using ${\cal Z}_2^{\,\citeloro}$ -- as obtained from
eq.~(\ref{eq:dZshift}) -- instead of ${\cal Z}_2^{\,\smalldr}$ and
then computing a non-minimal counterterm for $\tb$ via
eq.~(\ref{eq:dtanB}). We checked that, with this modification, we
reproduce exactly the corrections to the renormalized inverse
propagator shown in figures~5 and~10 of ref.~\cite{loro}. We also
reproduce the corrections to the scalar masses shown in figures~7, 8,
12 and~13 of that paper, although small discrepancies persist in the
case of the heaviest scalar when its mass is above the threshold
$\mH=2\,m_t$. These residual discrepancies are formally of higher
order in the perturbative expansion, and result from different
approximations in the two-step procedure for the determination of the
poles of the propagator (namely, we drop the imaginary parts of the
two-loop self-energies, while ref.~\cite{loro} keeps them).

In summary, we have found that in ref.~\cite{loro} the two-loop
renormalization of the Higgs fields and of the parameter $\tb$ is not
performed in the $\drbar$ scheme as claimed in the paper, but rather
in some non-minimal scheme where the WFR counterterms and $\delta \tb$
differ from their $\drbar$ counterparts by finite, non-polynomial
terms, and neither the Higgs fields nor $\tb$ obey their usual
renormalization-group equations (because of the explicit scale
dependence of the additional terms). This inconsistency should be
taken into account when comparing the results of ref.~\cite{loro} with
those of calculations that actually employ $\drbar$ definitions for
the WFR and for $\tb$. First of all, to account for the difference in
$\delta \tb$, the input value for the $\drbar$-renormalized parameter
$\tb$ should be converted to the corresponding value in the
non-minimal scheme of ref.~\cite{loro}, according to
\be
\label{shifttanB}
\tb^{\,\citeloro} ~=~ 
\tb^{\,\smalldr} ~-~ \frac{\at}{4\pi}\,N_c\,\tb~
\frac{\delta_\epsilon m_t}{m_t}~.
\ee
However, eqs.~(\ref{eq:dm11})--(\ref{eq:dm22}) show that the
contributions of $\delta \tb^{(2)}$ to the entries of the Higgs mass
matrix are suppressed by powers of $\cos\beta$. Consequently, the
effect on the Higgs masses arising from a two-loop difference in
$\delta \tb$ is very small already for $\tan\beta=5$. In fact, the
bulk of the numerical discrepancy between our results and those of
ref.~\cite{loro} is due to higher-order effects that are directly
related to the finite shift in the WFR. As discussed at the end of
section~\ref{sec:general}, such effects are included in the Higgs-mass
corrections when the latter are computed in terms of loop-corrected
Higgs masses and mixing, and can become numerically relevant when the
loop-corrected masses differ significantly from their tree-level
values.

\section{Conclusions}
\label{sec:concl}

We computed the two-loop corrections to the neutral MSSM Higgs masses
of $\oatas$ and $\oaas$ -- i.e., all two-loop corrections that involve
the strong gauge coupling when the only non-vanishing Yukawa coupling
is $h_t$ -- including the effect of non-vanishing external momenta in
the self-energies. We relied on an integration-by-parts technique to
express the momentum-dependent loop integrals in terms of a minimal
set of master integrals, and we used the public code {\tt TSIL} to
evaluate the latter. We obtained results for the Higgs-mass
corrections valid when all parameters in the one-loop part of the
corrections are renormalized in the $\drbar$ scheme, as well as
results valid in a mixed OS--$\drbar$ scheme where the top/stop
parameters are renormalized on-shell. Our results for the scalar and
pseudoscalar self-energies in the $\drbar$ scheme confirm the results
of an earlier calculation, ref.~\cite{martin_mom}, where they
overlap. In addition, we obtained new results for the two-loop
contributions to the $Z$-boson self-energy that involve the strong
gauge coupling. The latter, which were not computed in
ref.~\cite{martin_mom}, enter the $\oaas$ corrections to the Higgs
masses when the tree-level mass matrix of the scalars is expressed in
terms of the physical $Z$-boson mass.

We also compared our results for the momentum-dependent $\oatas$
corrections in the mixed OS--$\drbar$ scheme with those of a recent
calculation of the same corrections, ref.~\cite{loro}, and found
disagreement. We traced the reason for the discrepancy to the fact
that, contrary to what stated in ref.~\cite{loro}, in that calculation
the Higgs fields and the parameter $\tb$ are renormalized in a
non-minimal scheme instead of the usual $\drbar$ scheme. When this
difference is taken into account, we reproduce the results of
ref.~\cite{loro}, providing in passing a cross-check of the codes used
to evaluate the loop integrals in the two calculations (i.e., {\tt
  TSIL} and {\tt SecDec}, respectively). However, we noticed that {\tt
  TSIL}, which implements dedicated algorithms for two-loop
self-energy integrals, can be a thousand times faster than a
multi-purpose code such as {\tt SecDec} in the computation of the
Higgs-mass corrections. This should be taken into consideration when
including the momentum-dependent corrections in phenomenological
analyses of the MSSM parameter space.

As to the numerical impact of the corrections computed in this paper,
it could at best be described as moderate. We considered six benchmark
scenarios compatible with the results of Higgs and SUSY searches at
the LHC, and found that both the momentum-dependent part of the
$\oatas$ corrections and the whole $\oaas$ corrections can shift the
prediction for the lightest-scalar mass $\mh$ by several hundred
MeV. However, we noticed that -- at least in the considered scenarios
-- the two classes of corrections enter the prediction for $\mh$ with
opposite sign, and they compensate each other to a good extent. For
what concerns the heaviest-scalar mass $\mH$, the impact of the new
corrections is also modest, with the exception of regions around
real-particle thresholds where a fixed-order calculation is not
reliable anyway.

The predictions for the lightest-scalar mass, as obtained from popular
codes for the determination of the MSSM mass spectrum, carry a
theoretical uncertainty that has been estimated to be (at least) of
the order of $\pm 3$~GeV -- see, e.g.,
refs.~\cite{Degrassi:2002fi,Allanach:2004rh} and the more recent
discussion in ref.~\cite{Bagnaschi:2014rsa}. Against this backdrop,
the corrections presented in this paper can be considered
sub-dominant, and their inclusion in public codes might seem less
urgent than, e.g., the inclusion of the dominant three-loop
effects~\cite{3loopFD} or the proper resummation of large logarithms
in scenarios with multi-TeV stop
masses~\cite{heavystops,Bagnaschi:2014rsa}, both of which can shift
the prediction for the lightest-scalar mass by several
GeV. Nevertheless, one should not forget that the accuracy of the
measurement of the Higgs mass at the LHC has already reached the level
of a few hundred MeV -- i.e., comparable to our sub-dominant
corrections -- and will improve further when more data become
available. If SUSY shows up at last when the LHC operates at $13-14$
TeV, the Higgs mass will serve as a precision observable to constrain
MSSM parameters that might not be directly accessible by
experiment. To this purpose, the accuracy of the theoretical
prediction will have to match the experimental one, making a full
inclusion of the two-loop corrections to the Higgs masses
unavoidable. Our calculation should be regarded as a necessary step in
that direction.

\section*{Acknowledgments}
We thank S.P.~Martin and the authors of ref.~\cite{loro} for useful
discussions and for assistance in the comparison with the results of
their respective calculations.  
This work was supported in part by the Research Executive Agency
(REA) of the European Commission under the Grant Agreement number
PITN-GA-2010-264564 (LHCPhenoNet), and by the European Research
Council (ERC) under the Advanced Grant ERC-2012-ADG\_20120216-321133
(Higgs@LHC).
The work of P.S.~at LPTHE is supported in part by French state
funds managed by the ANR (ANR-11-IDEX-0004-02) in the context of the
ILP LABEX (ANR-10-LABX-63).

\vfill
\newpage

\def\bf{\rm}

\end{document}